# Light-driven C-H activation mediated by 2D transition metal dichalcogenides

*Jingang Li[1,2], Di Zhang[3], Zhongyuan Guo[3], Xi Jiang[4], Jonathan M. Larson[5], Haoyue Zhu[6], Tianyi Zhang[6], Yuqian Gu[7], Brian Blankenship[2], Min Chen[8], Zilong Wu[1], Suichu Huang[1], Robert Kostecki[9], Andrew M. Minor[8,10], Costas P. Grigoropoulos[2], Deji Akinwande[7], Mauricio Terrones[6,11,12], Joan M. Redwing[6,13], Hao Li[3\*], and Yuebing Zheng[1\*]*

[1] Materials Science & Engineering Program, Texas Materials Institute, and Walker Department of Mechanical Engineering, The University of Texas at Austin, Austin, TX 78712, USA

[2] Laser Thermal Laboratory, Department of Mechanical Engineering, University of California, Berkeley, CA 94720, USA

[3] Advanced Institute for Materials Research (WPI-AIMR), Tohoku University, Sendai 980-8577, Japan

[4] Materials Sciences Division, Lawrence Berkeley National Laboratory, Berkeley, CA 94720, USA

[5] Department of Chemistry and Biochemistry, Baylor University, Waco, TX 76798, USA

[6] Department of Materials Science and Engineering, The Pennsylvania State University, University Park, PA 16802, USA

[7] Chandra Family Department of Electrical & Computer Engineering, The University of Texas at Austin, Austin, TX 78712, USA

[8] Department of Materials Science and Engineering, University of California, Berkeley, CA 94720, USA

[9] Energy Storage and Distributed Resources Division, Lawrence Berkeley National Laboratory, Berkeley, CA 94720, USA

[10] National Center for Electron Microscopy, Molecular Foundry, Lawrence Berkeley National Laboratory, Berkeley, CA 94720, USA

[11] Center for Two-Dimensional and Layered Materials, The Pennsylvania State University, University Park, PA 16802, USA

[12] Department of Physics and Department of Chemistry, The Pennsylvania State University, University Park, PA 16802, USA

[13] 2D Crystal Consortium, Materials Research Institute, The Pennsylvania State University, University Park, PA 16802, USA

*Email: zheng@austin.utexas.edu (Y.Z.), li.hao.b8@tohoku.ac.jp (H.L.)

**Abstract**

**C-H bond activation enables the facile synthesis of new chemicals. While C-H activation in short-chain alkanes has been widely investigated, it remains largely unexplored for long-chain organic molecules. Here, we report light-driven C-H activation in complex organic materials mediated by 2D transition metal dichalcogenides (TMDCs) and the resultant solid-state synthesis of luminescent carbon dots in a spatially-resolved fashion. We unravel the efficient H adsorption and a lowered energy barrier of C-C coupling mediated by 2D TMDCs to promote C-H activation. Our results shed light on 2D materials for C-H activation in organic compounds for applications in organic chemistry, environmental remediation, and photonic materials.**

**Main text**

The emergence of C-H bond activation has provided revolutionary opportunities in organic chemistry, materials science, and biomedical engineering[1]. Specifically, the activation and functionalization of the ubiquitous C-H bonds enable new synthetic routes for functional molecules in a more straightforward and atom-economical way[2-5]. Since C-H bonds are thermodynamically strong and kinetically inert[6], many catalysts have been developed for C-H activation, including transition metals (e.g., palladium[7], cobalt[8], and gold[9,10]), zeolites[11,12], and metal-organic frameworks[13,14].

While intensive research efforts have been focused on C-H bonds in short-chain alkanes (e.g., methane and ethane)[15,16] and aromatic compounds[17], C-H activation in long-chain organic molecules is rarely reported. Yet, the derivation of C-H bonds in these complex molecules has significant potential in synthesizing functional organic complexes and transforming environmental pollutants (e.g., fossil-resource-derived hydrocarbons) into more valuable chemicals[18,19].

Herein, we report the light-driven C-H activation in long-chain molecules mediated by two-dimensional (2D) transition metal dichalcogenides (TMDCs). This TMDC-mediated C-H activation in organic molecules enables optical synthesis and patterning of luminescent carbon dots on solid substrates.

As a first example, we achieve the light-driven transformation of cetyltrimethylammonium chloride (CTAC, $C_{19}H_{42}ClN$), a long-chain quaternary ammonium surfactant[20], into luminescent carbon dots (CDs) on $WSe_2$ monolayers. By coupling experiments with density functional theory (DFT) calculations, we unravel the role of Se vacancies and oxidized states of $WSe_2$ in promoting the H adsorption. We further show that 2D TMDCs can facilitate the C-C coupling with a lowered energy barrier to catalyze C-H activation in complex organic molecules. This type of light-driven reaction mediated by 2D materials can be generalized to other long-chain organic compounds for the broader impacts on organic synthesis, chemical degradation, and photonics.

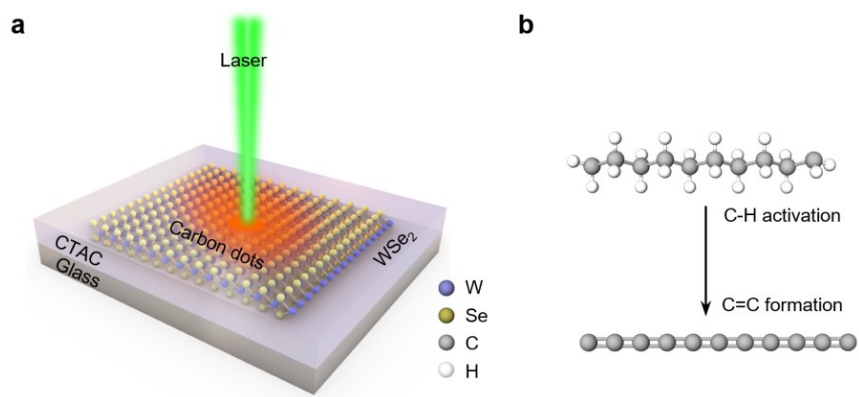

**Fig. 1. General concept of light-driven C-H activation in long-chain molecules mediated by 2D materials. a**, Schematic showing the light-driven transformation of CTAC on an atomic layer of $WSe_2$ into luminescent CDs. **b**, Schematic showing the photochemical reaction process involving the activation of C-H bonds and the formation of C=C bonds.

A typical experimental configuration is presented in Fig. 1a. A thin layer of solid CTAC is coated on a monolayer $WSe_2$ grown by chemical vapor deposition (CVD). The monolayer feature of $WSe_2$ is confirmed by the strong photoluminescence (PL) peak at ~ 750 nm (Fig. 2b, blue curve). Under the irradiation of a low-power continuous-wave laser (~ 0.2-5 mW), CTAC molecules undergo $WSe_2$-mediated C-H bond activation and the subsequent C=C bond formation (Fig. 1b). CTAC contains long carbon chains and quaternary ammonium cations, which has been commonly used as surfactants for chemical synthesis and fabric softeners[21]. Here, we choose CTAC as a first example due to its clean carbon-chain structure, solid form under ambient conditions, and wide existence in nanomaterials systems. This light-driven reaction can also be applied to other organic compounds.

The laser irradiation on hybrid CTAC/WSe$_2$ thin films leads to the emergence of bright luminescence from CDs (Fig. 2a). The evidence of CDs formation and materials characterizations are presented in Fig. 3. The optically generated CDs show pronounced broadband PL emission centered at ~ 600 nm under the excitation of a 532 nm laser (Fig. 2b, red curve). Additionally, the PL peak from WSe$_2$ exhibits a clear redshift from ~750 nm to ~780 nm, resulting from the charge transfer between the CDs and WSe$_2$[22,23]. Due to the negligible light absorption of CTAC and monolayer WSe$_2$ (Supplementary Fig. 1), we preclude the contribution of photothermal effects. Instead, this light-driven reaction is ascribed to the WSe$_2$-catalyzed C-H activation, which will be discussed later.

The photochemical reaction rate for the synthesis of CDs can be tuned by two orders of magnitude by controlling the laser power (Fig. 2c and Supplementary Movie 1). Under low-power laser irradiation, the emission of synthesized CDs remains stable for more than 20 min (Supplementary Fig. 2). Besides WSe$_2$, we also demonstrate the light-driven C-H activation and generation of CDs from CTAC on CVD-grown WS$_2$ and MoS$_2$ monolayers (Fig. 2d,e). Similar orangish PL emission from CDs can be directly visualized in optical imaging (Inset in Fig. 2d). The PL spectra of MoS$_2$/WS$_2$ + CDs also showed similar features, including a broadband emission from CDs centered at ~ 600 nm and a redshifted peak from MoS$_2$/WS$_2$. In addition, under the 660 nm laser excitation, the PL spectra from the WSe$_2$/WS$_2$ + CDs are distinct from those under the 532 nm excitation (Fig. 2f). This excitation wavelength-dependent PL emission is a characteristic feature of CDs[24,25].

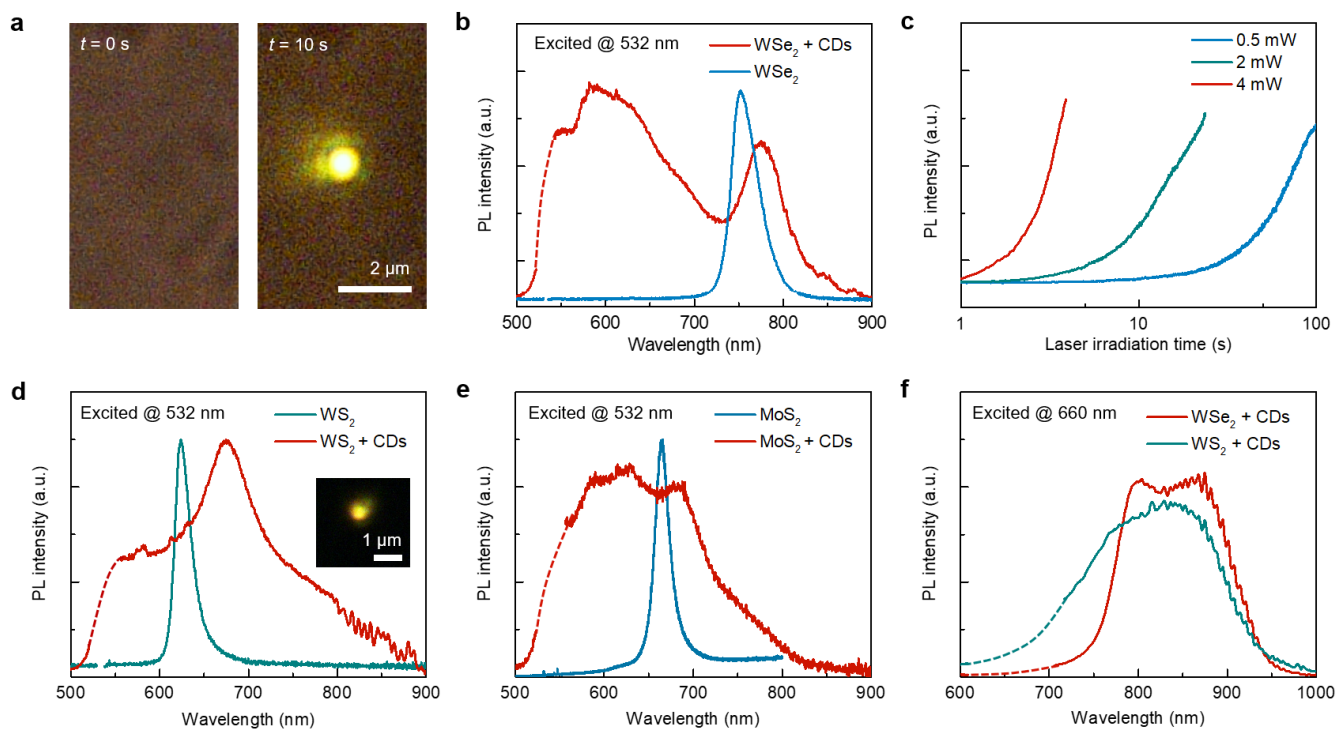

**Fig. 2. Optical characterizations of 2D-mediated C-H activation and CD synthesis. a**, Optical images showing the CTAC on the WSe$_2$ sample under a 532 nm laser irradiation at $t$ = 0 s and $t$ = 10 s. The laser power is 2.5 mW. The yellowish PL emission comes from the optically synthesized CDs. **b**, The PL spectra of WSe$_2$ and WSe$_2$ + CDs hybrids. **c**, Time-resolved PL intensity of CDs at 600 nm from the CTAC on WSe$_2$ sample under a 532 nm laser irradiation with different optical power. **d,e**, The PL spectra of (**d**) WS$_2$ and WS$_2$ + CDs hybrids and (**e**) MoS$_2$ and MoS$_2$ + CDs hybrids under the excitation of a 532 nm laser. Inset in (**d**): optical image showing the PL emission from the WS$_2$ + CDs sample. **f**, The PL spectra of WSe$_2$/WS$_2$ + CDs samples excited by a 660 nm laser.

The light-driven, 2D TMDC-mediated synthesis of CDs is confirmed by multiple characterization techniques. The Raman spectrum shows a D band at ~1380 cm$^{-1}$ and a G band at ~1600 cm$^{-1}$ (Fig. 3a), which are signatures of CDs[26]. The scanning electron microscope (SEM) images also reveal the existence of CD nanoparticles at the laser-irradiated areas (Fig. 3b,c). The as-synthesized CDs have a size distribution of 5-15 nm, as shown in the transmission electron microscope (TEM) images (Fig. 3e,f). The selected-area electron diffraction pattern exhibits bright diffraction spots and amorphous rings (Inset in Fig. 3f), indicating a semi-crystalline structure of CDs. The chemical composition of CDs is further examined by a near-field nanoscale Fourier transform infrared spectroscopy (nano-FTIR). Compared to the pristine CTAC film, the nano-FTIR spectrum of CDs presents a prominent absorption band at ~ 1660 cm$^{-1}$ (Fig. 3d), which is assigned to the vibrations of C=C bonds in CDs[27].

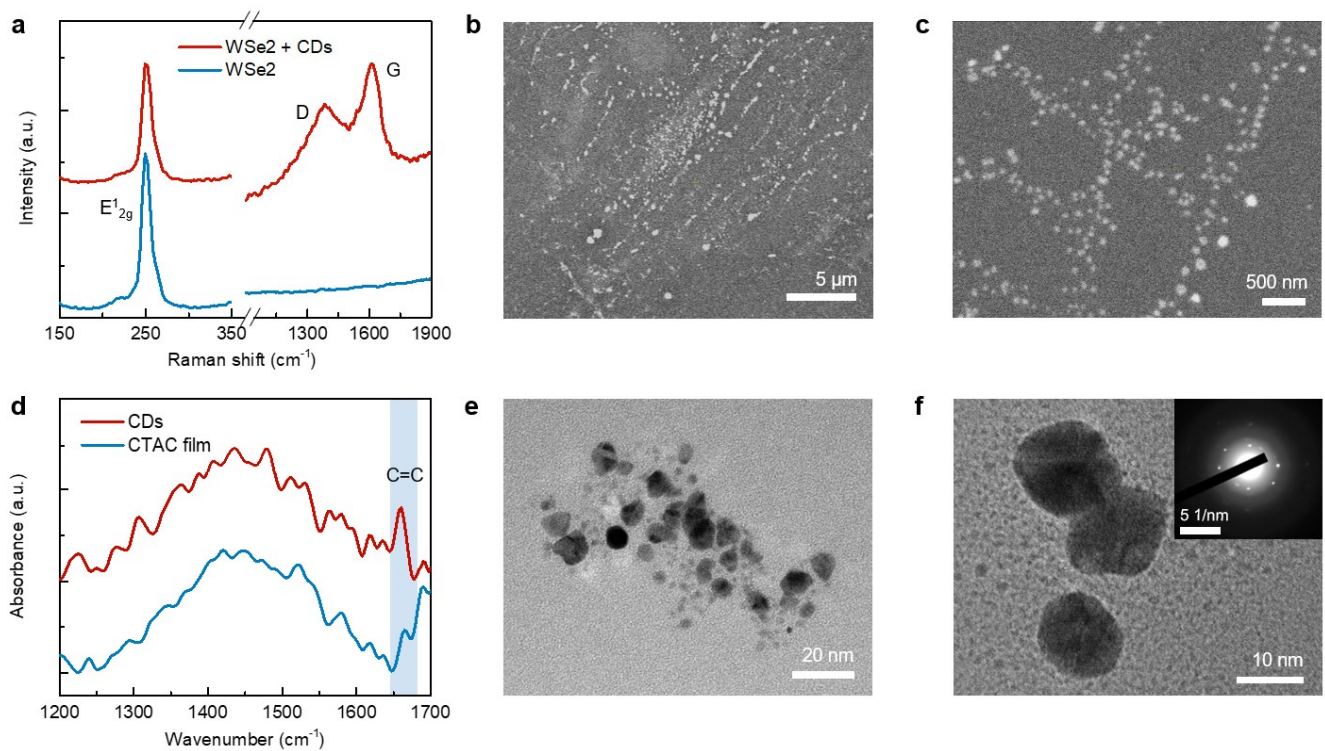

**Fig. 3. Materials characterizations of optically synthesized CDs. a**, Raman spectra of WSe$_2$ and WSe$_2$ + CDs hybrids. **b,c**, SEM images of the synthesized CDs. **d**, Near-field nano-FTIR spectra of the CDs and pristine CTAC films. **e,f**, High-resolution TEM images of the synthesized CDs. Inset in (**f**) shows the selected area electron diffraction (SAED) pattern of the CDs.

Next, we discuss the underlying mechanisms of the light-driven C-H bond activation medicated by 2D materials. C-H activation requires a sufficiently negative hydrogen adsorption free energy[28], however, pristine 2D TMDCs usually cannot meet this prerequisite since they are known to be facile hydrogen evolution materials[29]. To identify the potential active sites in our study that drive the C-H bond activation, we first measured the X-ray photoelectron spectroscopy spectra of the monolayer WSe$_2$. The results indicate the existence of prevalent Se vacancies and O adsorption on the CVD-grown WSe$_2$ surfaces (Supplementary Fig. 3)[30,31]. To analyze the role of Se vacancies and O substitution on WSe$_2$, we calculated the projected density of states (PDOS) of local W-sites using DFT calculations (Fig. 4a and Supplementary Fig. 4). With the increasing number of Se vacancies, there is an obvious shift of the peak toward the Fermi level (Fig. 4b). The calculated average energies of the $d$-electrons (i.e., the $d$-band center) of the sites with Se vacancies are also closer to the Fermi level compared to a pristine $WSe_2$. According to the $d$-band center theory[32], a surface site with a $d$-band center closer to the Fermi level corresponds to a significantly stronger H adsorption capacity[33], which facilitates the C-H bond activation due to the stronger driving force to "pull" a H down to the surface[34]. Similar conclusions can be found on a $WSe_2$ surface with oxygen substitution at Se sites (Fig. 4c). Meanwhile, the existence of adsorbed oxygen and the subsequently formed hydroxyl can act as the promoters to expedite C-H activation due to a facile O/HO-promoted mechanism[35-38]. To verify the theoretical hypothesis, we conducted control experiments on mechanically exfoliated $WSe_2$ flakes with fewer surface defects[39], and the results show that a much higher optical power is required for this reaction to occur (Supplementary Fig. 5). These theoretical analyses and experiments indicate that the Se vacancy and O substitution in $WSe_2$ can both lead to a more facile C-H activation capacity due to either higher reactivity of a defected surface or an O-promotion effect.

For long carbon chains, the C-H activation is followed by the formation of C=C bonds[40]. We further investigate the capability of 2D TMDCs to drive the C=C formation. We analyze the C-C coupling on material surfaces (Fig. 4d), where two carbon atoms are bonded together. We compare the calculated kinetic energy barriers of this process for $WSe_2$ and other common catalyst surfaces for C-H activation (Supplementary Fig. 6), including gold (Au) and palladium (Pd). The energy barrier of C-C coupling on $WSe_2$ surfaces is calculated to be 0.29 eV (Fig. 4e), which is significantly lower than Au (0.57 eV) and Pd (1.29 eV). These results indicate that while metal catalysts (e.g., Pd and Au) are suitable for C-H activation in short-chain molecules, they cannot be generalized to long carbon chains due to the high activation energy of C-C coupling to form C=C bonds. This energy barrier is further reduced to 0.23 eV on $WSe_2$ surfaces with Se vacancies (Fig. 4e and Supplementary Fig. 7). These results demonstrate the potential of

2D WSe$_2$ as promising catalysts to drive the C-H activation of long-chain molecules and facilitate the subsequent C=C formation.

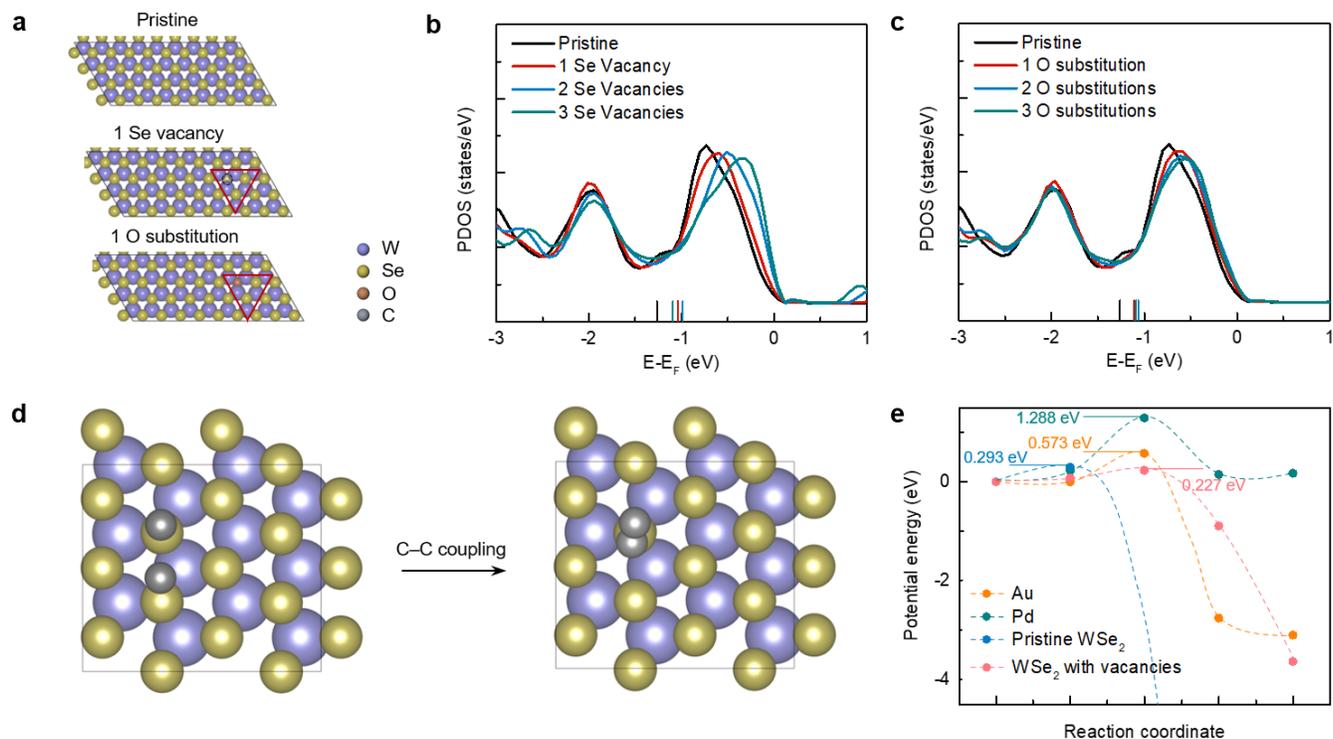

**Fig. 4. First-principles calculations to provide insights into the light-driven C-H activation mediated by 2D materials. a**, Optimized structures considered for DFT calculations. Pristine WSe$_2$ and WSe$_{2-x}$ with Se vacancies or O substitutions are considered. **b,c**, PDOS of the *d*-electrons of local W-sites (red triangles in c) at pristine WSe$_2$ and WSe$_{2-x}$ with Se vacancies (**b**) or O substitutions (**c**). The calculated *d*-band centers are marked with vertical lines. The Fermi levels are shifted to zero. **d**, The process of C-C coupling considered for DFT calculations on the WSe$_2$ surface. **e**, Comparison of the kinetic barriers of C-C coupling on the WSe$_2$ and other surfaces.

In summary, we discover the 2D-TMDC-mediated C-H activation in long-chain organic molecules under light illumination. Our experimental characterizations coupled with theoretical calculations reveal the role of defects and oxidized states on TMDCs in the promotion of H adsorption and C-H activation reactions. Moreover, we find that the energy barrier of C-C coupling mediated by 2D TMDCs is much lower than the commonly used metal catalysts for C-H activation of short-chain alkanes, highlighting its promising performance of C-H activation for complex molecules.

This light-controlled site-specific C-H activation also enables optical printing of luminescent carbon dots on solid substrates and provides an approach towards data encryption and information technology[41]. By controlling the thickness of CTAC layer, laser power, and irradiation time, we can write CDs by laser scanning without changing the morphology of the film (Supplementary Fig. 8). Thus, the embedded patterns remain hidden under white light illumination and can be read out by fluorescence, Raman, or PL imaging (Supplementary Fig. 9). Besides CTAC, this strategy is general to other long-chain molecules, such as octyltrimethylammonium chloride and polyvinyl alcohol (Supplementary Fig. 10). We envision that the 2D-TMDC-mediated light-driven C-H activation in complex organic molecules will enable new applications in chemical synthesis, photonics, degradation of organic pollutants, and plastic recycling.

## Methods

**Chemicals and materials.** CTAC was purchased from Chem-Impex. Other chemicals, including octyltrimethylammonium chloride and polyvinyl alcohol, were purchased from Sigma-Aldrich. All the materials were used without further purification. The CVD-grown monolayer $WSe_2$, $WS_2$, and $MoS_2$ were synthesized based on the methods that have been described in detail in previous reports (ref.[42] for $WSe_2$, ref.[43] for $WS_2$, and ref.[44] for $MoS_2$).

**Optical setup.** The light-driven C-H activation and laser writing of CDs were performed in a Nikon inverted microscope (Nikon TiE) equipped with a ×100 oil objective (Nikon, NA 0.5–1.3), a halogen white light source (12 V, 100 W), a bright-field or dark-field condenser (NA 1.20–1.43), and a color charge-coupled device (CCD) camera (Nikon). A 532 nm laser (Coherent, Genesis MX STM-1 W) or a 660 nm laser (Laser Quantum) was expanded with a 5× beam expander (Thorlabs, GBE05-A) and directed to the microscope.

**Characterizations.** The Raman spectra and mapping were measured on a Renishaw system using a 532 nm wavelength laser source. The absorption spectra and PL spectra were recorded with a spectrograph (Andor) and an EMCCD (Andor) integrated into an inverted optical microscope. The scanning electron microscopy (SEM) images were taken with a FEI Quanta 650 SEM. TEM images and diffraction patterns were obtained with a JEOL 1400 (120kV) with Gatan Inc. One view camera and a specialized TEM holder (Laser Prismatics). Near-field nano-FTIR measurements were performed with a commercial Neaspec system equipped with a broadband laser source[45]. The XPS spectra were collected on a Kratos AXIS Ultra XPS spectrometer.

**DFT calculations.** All DFT calculations were performed using the VASP code with the valence electrons treated by expanding the Kohn-Sham wave functions in a planewave basis set[46]. The method of generalized gradient approximation using the Revised Perdew-Burke-Ernzerhof (RPBE) functional was employed to describe the electronic exchange and correlations[47]. The core electrons were treated by the projector augmented wave method[48]. Van der Waals corrections were included within Grimme's framework (DFT+D3)[49]. Convergence was defined when the forces of each atom fell below 0.05 eV per Å. The energy cutoff was set to 400 eV. A (3×3×1) $k$ point mesh was employed to sample the Brillouin zone based on the method of Monkhorst and Pack[50]. The kinetic barriers were calculated based on the climbing-image nudged elastic band (CI-NEB) method[51]. To ensure sufficient spacing, we placed a vacuum spacing of at least 12 Å perpendicular to the surface.

## Data availability

All data that support the findings of this study are included in the paper and/or Supplementary Materials.

## Acknowledgements

Y.Z. acknowledges the financial support of the National Institute of General Medical Sciences of the National Institutes of Health (DP2GM128446) and the National Science Foundation (NSF-ECCS-2001650). H. L. acknowledges the JSPS KAKENHI (Grant No. JP23K13703), the Iwatani Naoji Foundation. J.L. acknowledges the financial support of the University Graduate Continuing Fellowship from The University of Texas at Austin. J.M.R and H.Z. acknowledge the financial support of the National Science Foundation through the Penn State 2D Crystal Consortium - Materials Innovation Platform (2DCC-MIP) under NSF cooperative agreement DMR-2039351. B.B acknowledges support from the NSF Graduate Research Fellowship (DGE 2146752). TEM imaging work was funded by the Soft Matter Electron Microscopy Program (KC11BN), supported by the Office of Science, Office of Basic Energy Science, US Department of Energy, under Contract DE-AC02-05CH11231. The authors acknowledge

**Author contributions**

J.L., Y.Z. and H.L. conceived the idea and designed the research. J.L. prepared the samples, worked on the experiments, and collected the data. H.L., D.Z. and Z.G. worked on the DFT calculations. X.J., B.B., M.C., A.M.M. and C.P.G. worked on TEM imaging. J.M.L. performed the nano-FTIR measurement. H.Z. and J.M.R. synthesized $WSe_2$. T.Z. and M.T. synthesized $WS_2$. Y.G. synthesized $MoS_2$ under the supervision of D.A.. Z.W. and S.H. assisted in sample preparation and experiments. Y.Z. supervised the project. J.L., H.L. and Y.Z. wrote the manuscript with inputs from all authors.

**Competing interests**

The authors declare that they have no competing interests.

Supplementary Materials for

# Light-driven C-H activation mediated by 2D transition metal dichalcogenides

*Jingang Li[1,2], Di Zhang[3], Zhongyuan Guo[3], Xi Jiang[4], Jonathan M. Larson[5], Haoyue Zhu[6], Tianyi Zhang[6], Yuqian Gu[7], Brian Blankenship[2], Min Chen[8], Zilong Wu[1], Suichu Huang[1], Robert Kostecki[9], Andrew M. Minor[8,10], Costas P. Grigoropoulos[2], Deji Akinwande[7], Mauricio Terrones[6,11,12], Joan M. Redwing[6,13], Hao Li[3\*], and Yuebing Zheng[1\*]*

[1] Materials Science & Engineering Program, Texas Materials Institute, and Walker Department of Mechanical Engineering, The University of Texas at Austin, Austin, TX 78712, USA

[2] Laser Thermal Laboratory, Department of Mechanical Engineering, University of California, Berkeley, CA 94720, USA

[3] Advanced Institute for Materials Research (WPI-AIMR), Tohoku University, Sendai 980-8577, Japan

[4] Materials Sciences Division, Lawrence Berkeley National Laboratory, Berkeley, CA 94720, USA

[5] Department of Chemistry and Biochemistry, Baylor University, Waco, TX 76798, USA

[6] Department of Materials Science and Engineering, The Pennsylvania State University, University Park, PA 16802, USA

[7] Chandra Family Department of Electrical & Computer Engineering, The University of Texas at Austin, Austin, TX 78712, USA

[8] Department of Materials Science and Engineering, University of California, Berkeley, CA 94720, USA

[9] Energy Storage and Distributed Resources Division, Lawrence Berkeley National Laboratory, Berkeley, CA 94720, USA

[10] National Center for Electron Microscopy, Molecular Foundry, Lawrence Berkeley National Laboratory, Berkeley, CA 94720, USA

[11] Center for Two-Dimensional and Layered Materials, The Pennsylvania State University, University Park, PA 16802, USA

[12] Department of Physics and Department of Chemistry, The Pennsylvania State University, University Park, PA 16802, USA

[13] 2D Crystal Consortium, Materials Research Institute, The Pennsylvania State University, University Park, PA 16802, USA

*Email: zheng@austin.utexas.edu (Y.Z.), li.hao.b8@tohoku.ac.jp (H.L.)

**This PDF file includes**

    Supplementary Figs. 1-10

**Other Supplementary Materials include**

    Supplementary Movie 1: PL emission from CTAC on $WSe_2$ sample under different laser power

**Supplementary Figures**

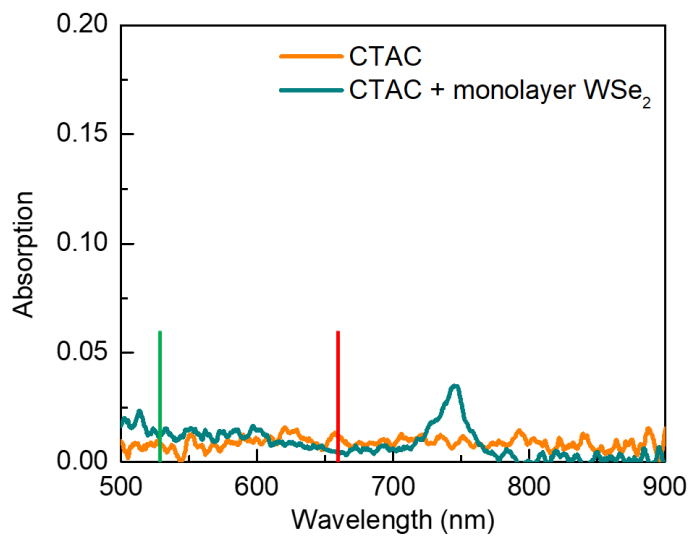

**Supplementary Fig. 1. Measured optical absorption of CTAC and monolayer WSe₂.** The green and red vertical lines indicate the laser wavelengths used in this work, 532 nm and 660 nm, respectively.

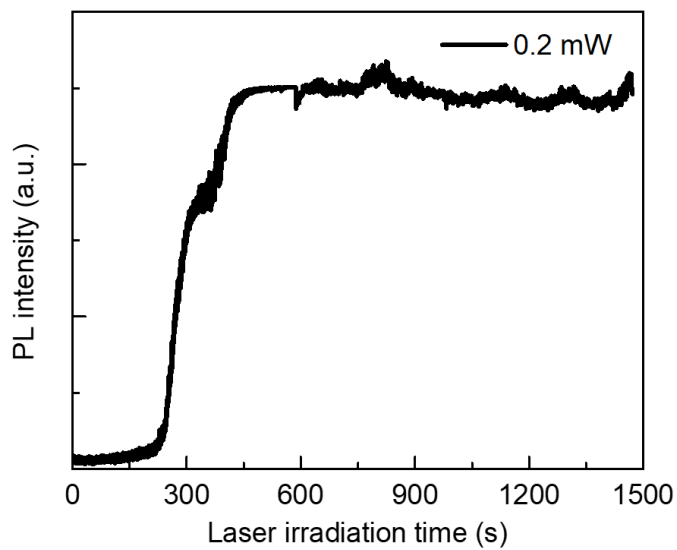

**Supplementary Fig. 2. Time-resolved PL intensity of CDs at 600 nm from the CTAC on WSe₂ sample under a 532 nm laser irradiation.** The optical power is 0.2 mW.

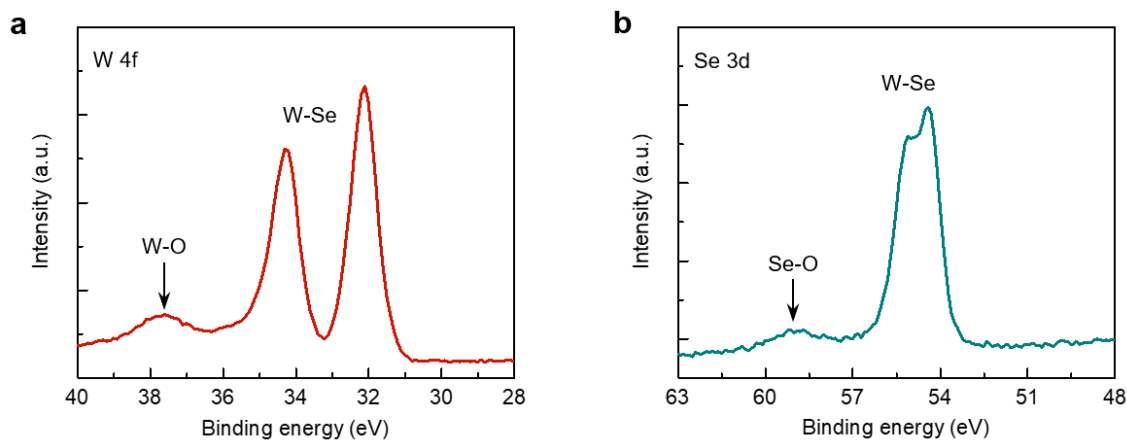

**Supplementary Fig. 3. High-resolution X-ray photoelectron spectroscopy (XPS) spectra of W 4f (a) and Se 3d (b) regions of CVD-grown monolayer WSe$_2$.** In addition to the peaks of W $4f_{7/2}$ (~32 eV) and W $4f_{5/2}$ (~34.1 eV) of WSe$_2$, a small peak at ~37.5 eV can be observed, which corresponds to the W $4f_{5/2}$ from WO$_3$. Similarly, the appearance of a small peak at ~59 eV for Se 3d suggests the formation of Se-O bonding. The Se vacancies and O adsorption on the surfaces have been regarded as ubiquitous for CVD-grown 2D WSe$_2$.

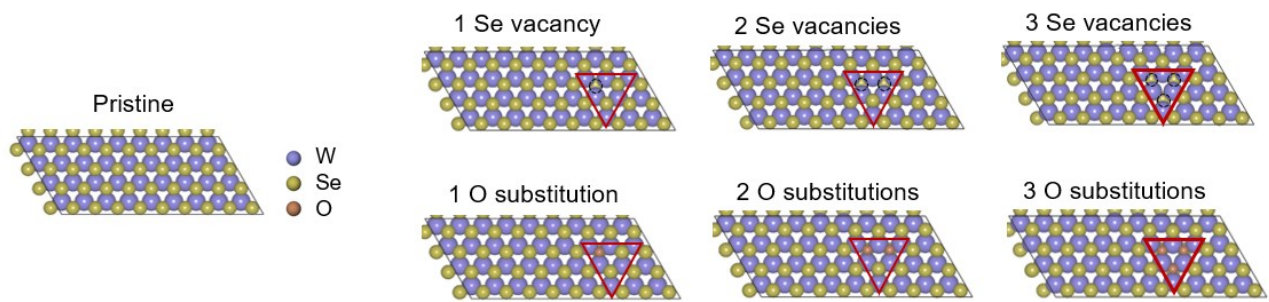

**Supplementary Fig. 4. Optimized structures considered for DFT calculations in Fig. 4.** Pristine WSe$_2$ and WSe$_{2-x}$ with Se vacancies or O substitutions are considered.

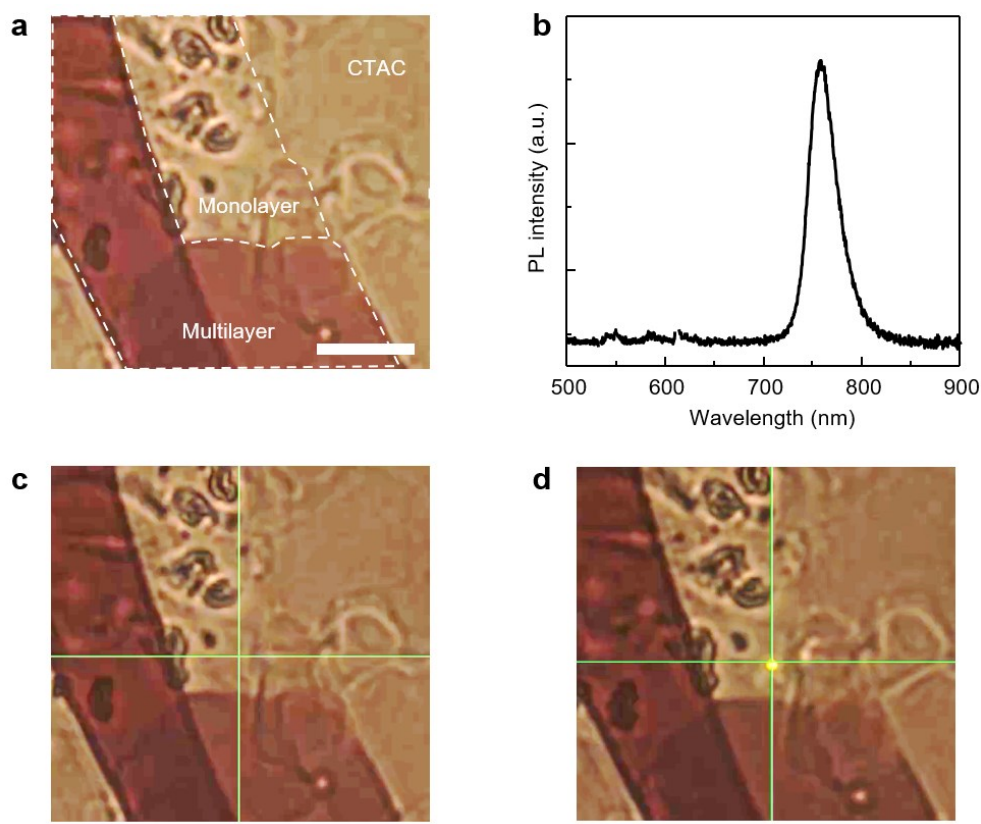

**Supplementary Fig. 5. Light-driven C-H activation and carbon dots formation on mechanically exfoliated WSe₂. a**, Optical image showing the exfoliated WSe$_2$ flake coated with a thin layer of CTAC. Scale bar: 10 μm. **b**, Measured photoluminescence (PL) spectrum at the monolayer region. The strong PL mission confirms the monolayer feature. **c**, No obvious PL emission from carbon dots after laser irradiation at 5 mW for 5 min. **d**, The evolution of carbon dots and PL after laser irradiation at 15 mW for 1 min. These results show that a much higher optical power is required for this reaction to occur on exfoliated WSe$_2$ flake than that on CVD-grown monolayer WSe$_2$, where the lowest power can be down to 0.2 mW.

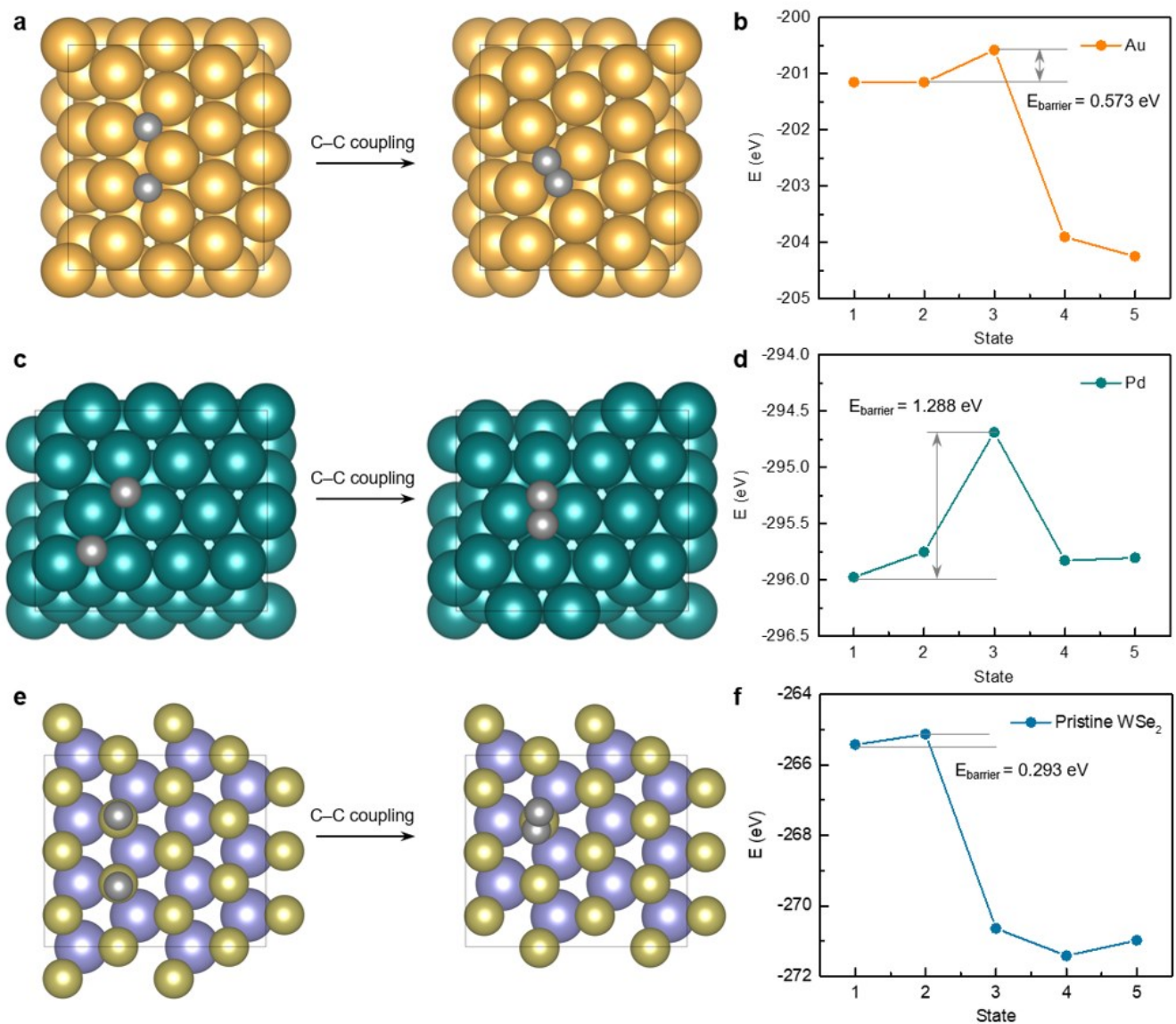

**Supplementary Fig. 6. DFT calculations of C-C coupling on different surfaces. a,c,e,** Initial and final states of C-C coupling on (**a**) gold, (**c**) palladium, and (**e**) pristine WSe$_2$ surfaces. **b,d,f,** The energy evolution during the C-C coupling on (**b**) gold, (**d**) palladium, and (**f**) pristine WSe$_2$ surfaces. "1" and "5" denote the initial and final states, respectively. "2-4" are intermediate states.

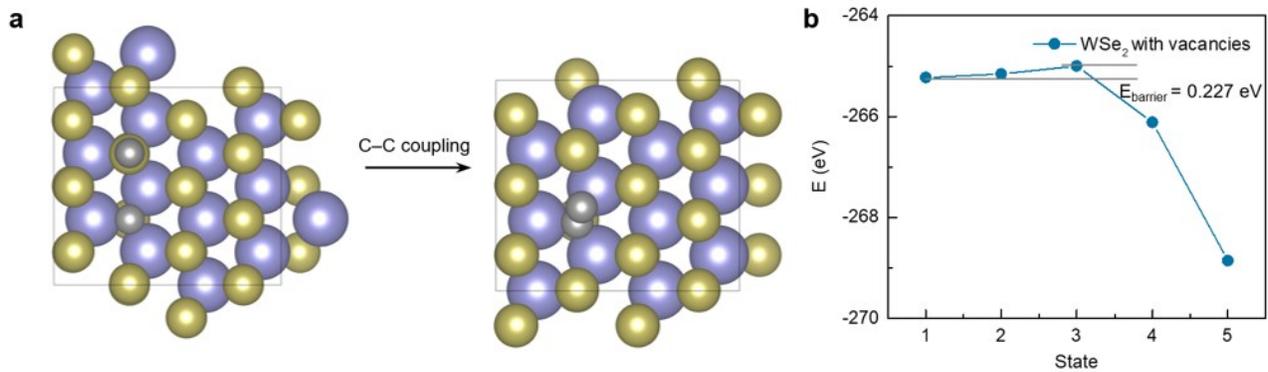

**Supplementary Fig. 7. DFT calculations of C-C coupling on WSe₂ surfaces with Se vacancies. (a)** Initial and final states and **(b)** the energy evolution during the C-C coupling. "1" and "5" denote the initial and final states, respectively. "2-4" are intermediate states.

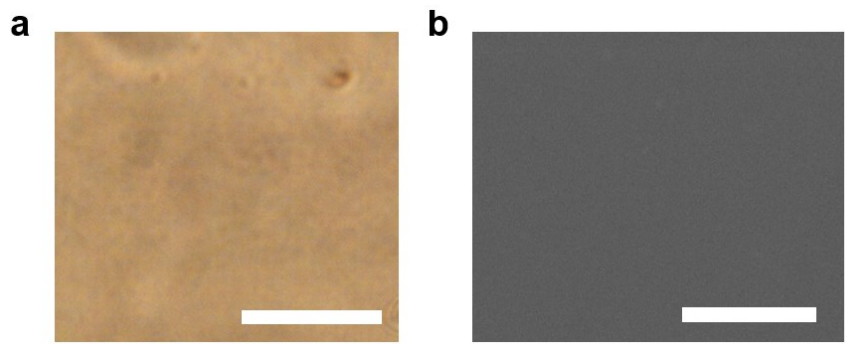

**Supplementary Fig. 8. Characterization of the CTAC layer after the laser writing of carbon dots.** **(a)** Optical image and **(b)** scanning electron microscopic image of the CTAC layer after the laser writing of carbon dots. The optical power used is 2 mW; the laser wavelength is 532 nm; the laser irradiation time is approximately 10 seconds. Scale bars: 10 μm.

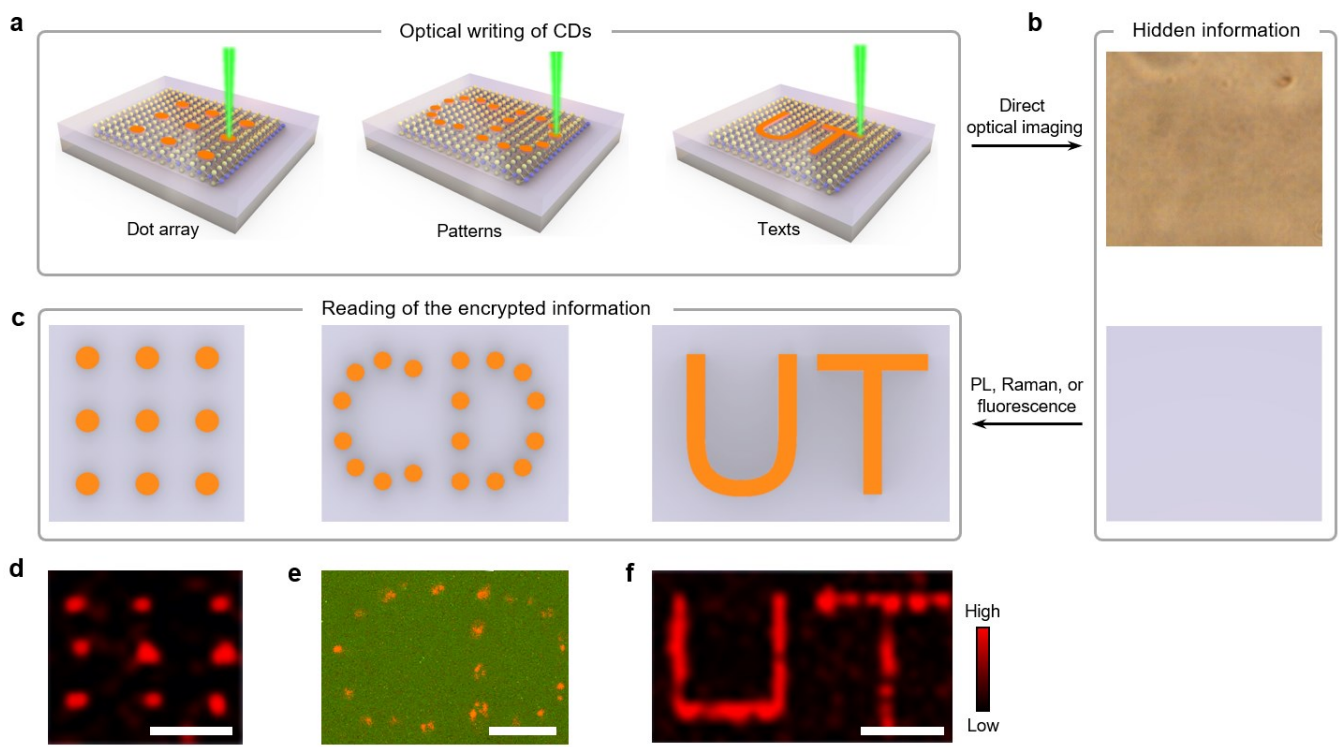

**Supplementary Fig. 9. Optically generated CDs for information encryption. a**, Schematic showing the optical writing of CDs. **b**, The written CDs remain hidden under direct bright-field optical imaging. **c**, Schematic showing the read-out of the encrypted information by PL, Raman, or fluorescence imaging. **d-f**, PL mapping (**d**), fluorescent imaging (**e**), and Raman mapping (**f**) mapping of the encrypted CDs patterns. All scale bars are 10 μm.

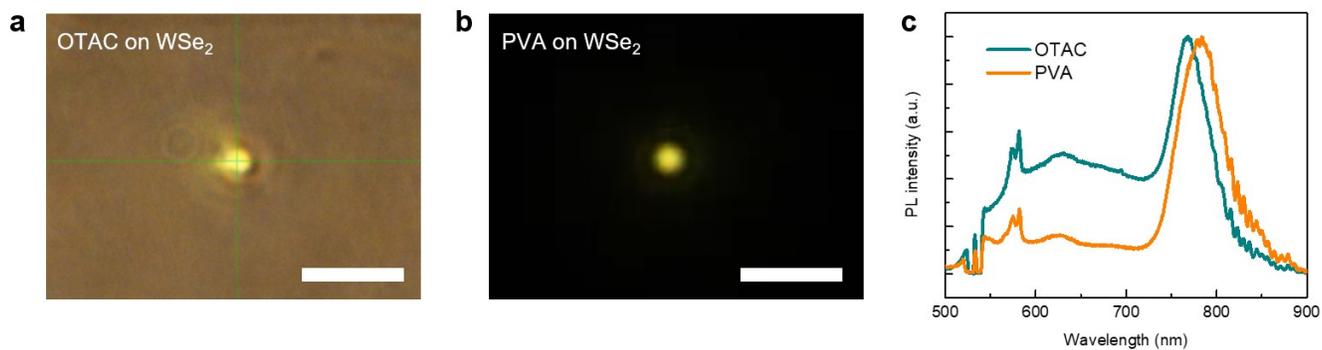

**Supplementary Fig. 10. Light-driven C-H activation and carbon dots formation from other organic molecules on a monolayer WSe₂. a,b**, Optical images showing the PL emission of carbon dots from (**a**) octyltrimethylammonium chloride (OTAC) and (**b**) polyvinyl alcohol (PVA) on monolayer $WSe_2$. The laser wavelength is 532 nm. Scale bars: 10 μm. **c**, Measured PL spectra from OTAC and PVA on $WSe_2$ sample, both showing the emission bands from CDs.